\documentclass[twocolumn,letterpaper,aps,prc,superscriptaddress,showpacs,floatfix,nofootinbib]{revtex4}

\usepackage{graphicx}	
\usepackage{xspace}     
\usepackage{amsmath}
\usepackage[usenames,dvipsnames]{color}
\usepackage{hyperref}

\newcommand{\sqsn}{\mbox{$\sqrt{s_{_{NN}}}$}\xspace}

\def\lsim{\raise0.3ex\hbox{$<$\kern-0.75em\raise-1.1ex\hbox{$\sim$}}}
\def\gsim{\raise0.3ex\hbox{$>$\kern-0.75em\raise-1.1ex\hbox{$\sim$}}}
\def\mean#1{\left<#1\right>}
\def\Journal#1#2#3#4{{#1}{\bf #2}, #3 (#4)}

\def\EPJC{{Eur. Phys. J. C}}

\def\JPCS{{J. Phys: Conf. Series\ }}

\def\NPB{{Nucl. Phys. B}}
\def\PLB{{Phys. Lett. B}}
\def\PLC{Phys. Repts.\ }

\def\PRL{Phys. Rev. Lett.\ }
\def\PRD{{Phys. Rev. D}}
\def\PRC{{Phys. Rev. C}}

\def\ZPC{{Z. Phys. C}}

\begin{document}
\title{Improved Comparison of Measurements and Calculations of $\hat{q}L$ via transverse momentum broadening in Relativistic Heavy Ion Collisions using di-hadron correlations.}

\author{M.~J.~Tannenbaum\\ {\it Brookhaven National Laboratory, Upton, NY 11973 USA}}

\date{\today}

\begin{abstract}
The renewed interest in analyzing RHIC data on di-hadron correlations as probes of final state transverse momentum broadening as shown at Quark Matter 2018~\cite{MiklosQM2018} by theoretical calculations~\cite{ChenCCNUPLB773} compared to experimental measurements~\cite{PXppg074, STARPLB760} led me to review the quoted theoretical calculations and experimental measurements because the theoretical calculation~\cite{ChenCCNUPLB773} does not show the PHENIX measurements~\cite{PXppg074} as published. The above references were checked and fits were performed to the published measurements~\cite{PXppg074,PXppg083} to determine $\hat{q}L$ from the measured azimuthal broadening to compare with the theoretical calculation~\cite{ChenCCNUPLB773}. The new results will be presented in addition to some corrections to the previous work~\cite{MJTPLB771}. The measured values of $\hat{q}L$ show the interesting effect of being consistent with zero for larger values of associated $p_{Ta}\geq3$ GeV/c which is  shown to be related to well known measurements of the ratio of the Au+Au to p+p associated $p_{Ta}$ distributions for a given trigger $p_{Tt}$ called $I_{AA}$~\cite{ppg106,ALICEPLB763}. Di-jets rather than di-hadrons are proposed as an improved azimuthal broadening measurement to determine $\hat{q}L$ and possibly $\hat{q}$.
\end{abstract}

	
\maketitle

\section{Introduction} 
\begin{figure*}[!h]
\begin{center}
\includegraphics[width=0.9\linewidth]{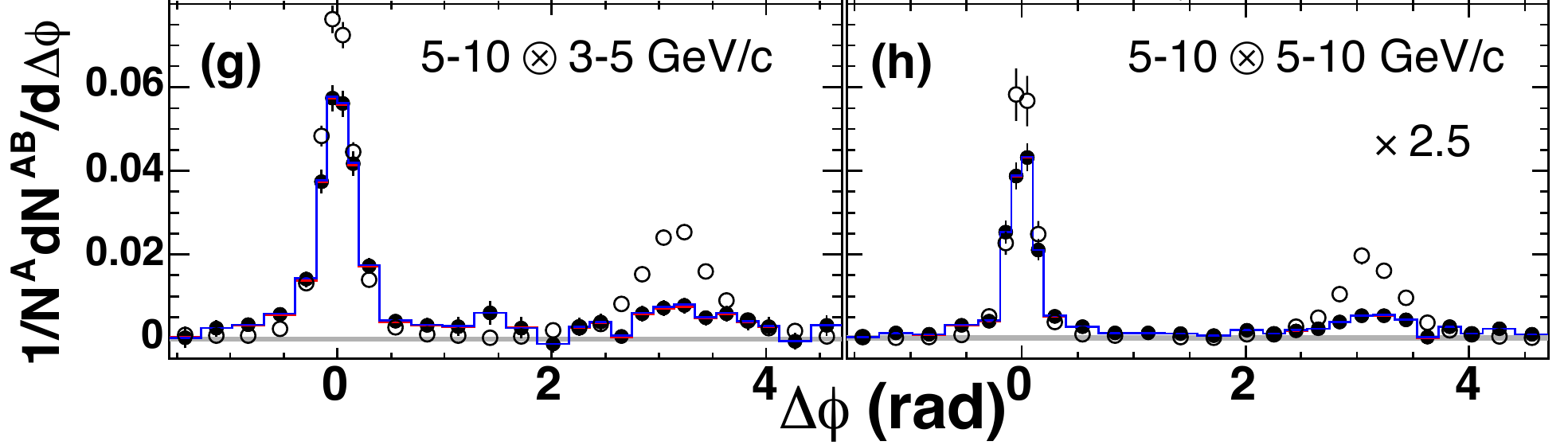}
\end{center}
\caption[]{PHENIX azimuthal correlation conditional yield in p$+$p and 0-20\% centrality Au$+$Au collisions at \sqsn=200 GeV for trigger $h^{\pm}$ with $p_{Tt}=5 - 10$ GeV/c and associated $h^{\pm}$ with  $p_{Ta}=3-5$ GeV/c (g) and  $p_{Ta}=5-10$ GeV/c (h) reproduced from Ref.~\cite{PXppg074}} 
\label{fig:ppg074gh}
\end{figure*}

\begin{figure*}[!hbt]
\begin{center}
\includegraphics[width=5.55cm]{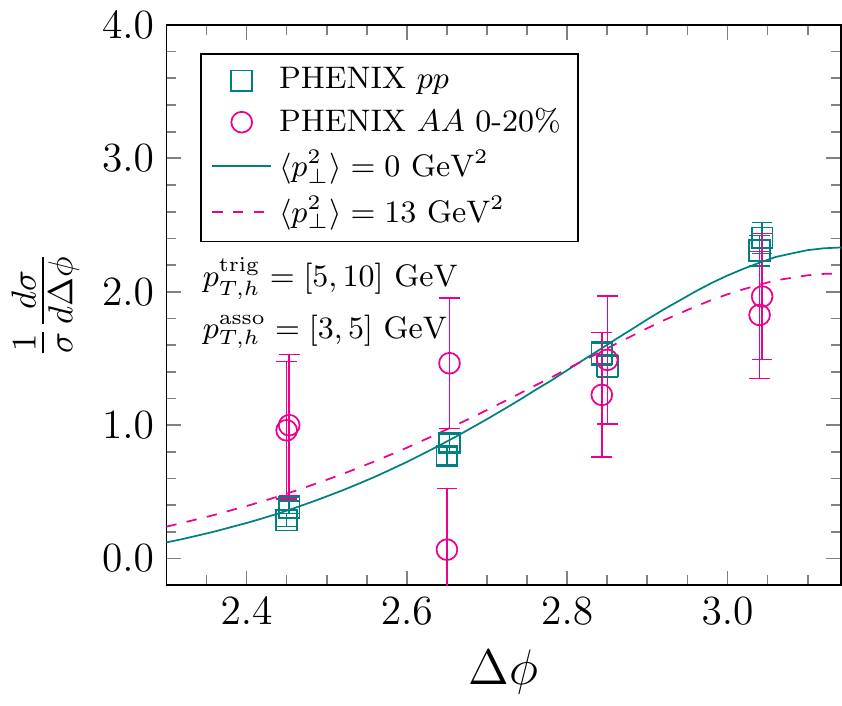}
\includegraphics[width=5.55cm]{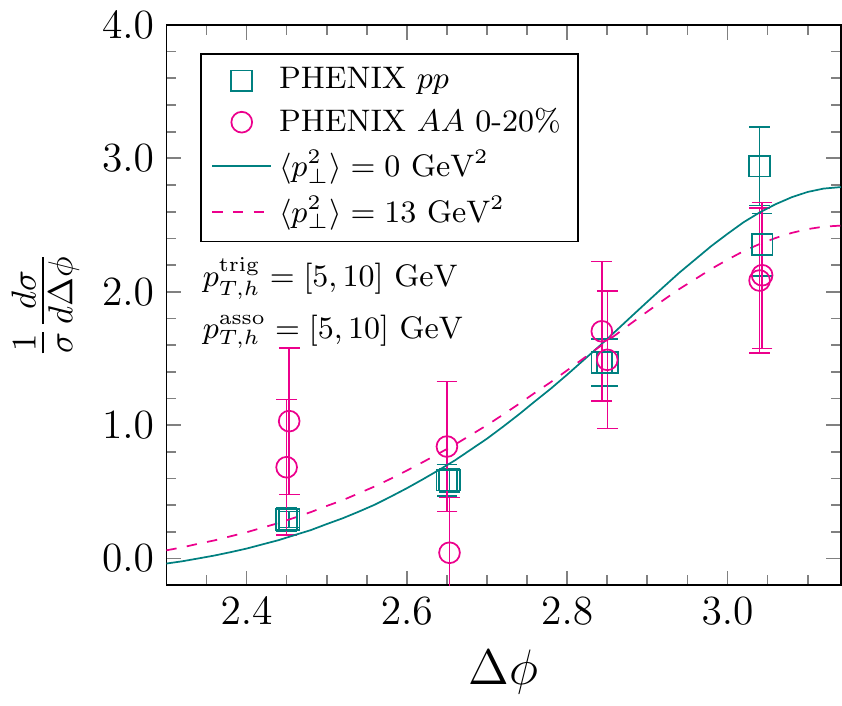}
\includegraphics[width=5.55cm]{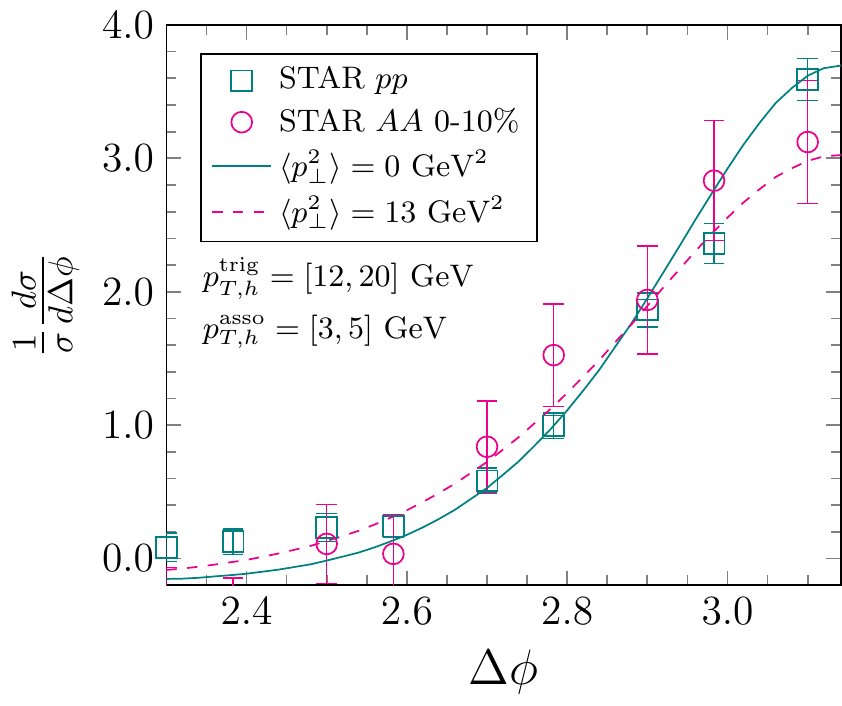}
\end{center}\vspace{-5.0pt}
\caption[*]{Figure of Normalized dihadron angular correlation compared with PHENIX \cite{PXppg074} and STAR \cite{STARPLB760} data, reproduced from Ref.~\cite{ChenCCNUPLB773}}
\label{fig:Bowendihadron}
\end{figure*}

  When I was reviewing talks from Quark Matter 2018, a slide in a presentation by Miklos Gyulassy~\cite{MiklosQM2018} drew my attention because it involved a figure (Fig.~\ref{fig:Bowendihadron}) from a preprint~\cite{ChenCCNU0616} that I had referenced in my publication on measuring $\hat{q}L$ from di-hadron correlations~\cite{MJTPLB771}. I had not paid much attention to that figure previously, but in comparing the two plots labeled PHENIX in Fig.~2 which is reproduced from Ref.~\cite{ChenCCNUPLB773} to the actual plots in Fig.~1 reproduced from the quoted PHENIX publication~\cite{PXppg074} I realized that the the data in the plots in Fig.~2 reproduced from Ref.~\cite{ChenCCNUPLB773} looked nothing like the measurement shown in Fig.~\ref{fig:ppg074gh} reproduced from the quoted PHENIX publication~\cite{PXppg074}. Notably, in the actual PHENIX data~\cite{PXppg074} shown in Fig.~1 errors are shown for the same side peaks in p$+$p and Au$+$Au, but no errors are shown for the away-side peaks ($\pi/2<\Delta\phi<3\pi/2$ radians) for either p$+$p or Au$+$Au. However, in Fig.~2 reproduced from Ref.\cite{ChenCCNUPLB773} which is labelled as PHENIX data from reference~\cite{PXppg074} errors are shown for the p$+$p and Au$+$Au away-side data.  
  
To understand this issue, I checked with  the authors of Fig.~\ref{fig:Bowendihadron} reproduced from Ref.~\cite{ChenCCNUPLB773} who informed me that software called xyscan was used on Figure~1 reproduced from Ref.~\cite{PXppg074} to get the data points and the error used in Fig.~2 reproduced from Ref.~\cite{ChenCCNUPLB773}. Also the the points for both p$+$p and A$+$A data were rescaled to make them normalize to 1 in Ref.\cite{ChenCCNUPLB773}, which were called `self normalized' data.

\begin{figure*}[!hbt]
\begin{center}
\includegraphics[width=5.55cm]{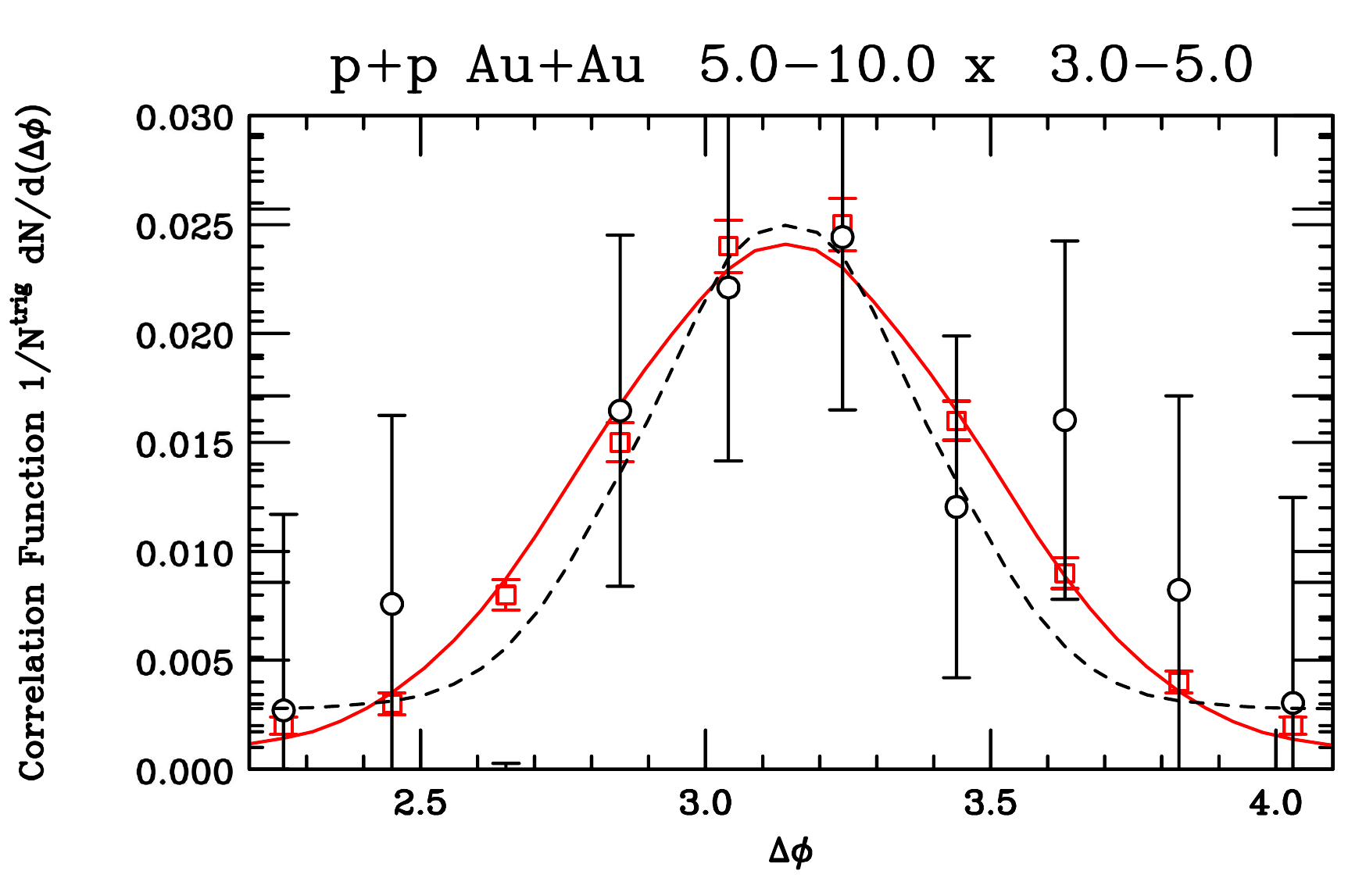}
\includegraphics[width=5.55cm]{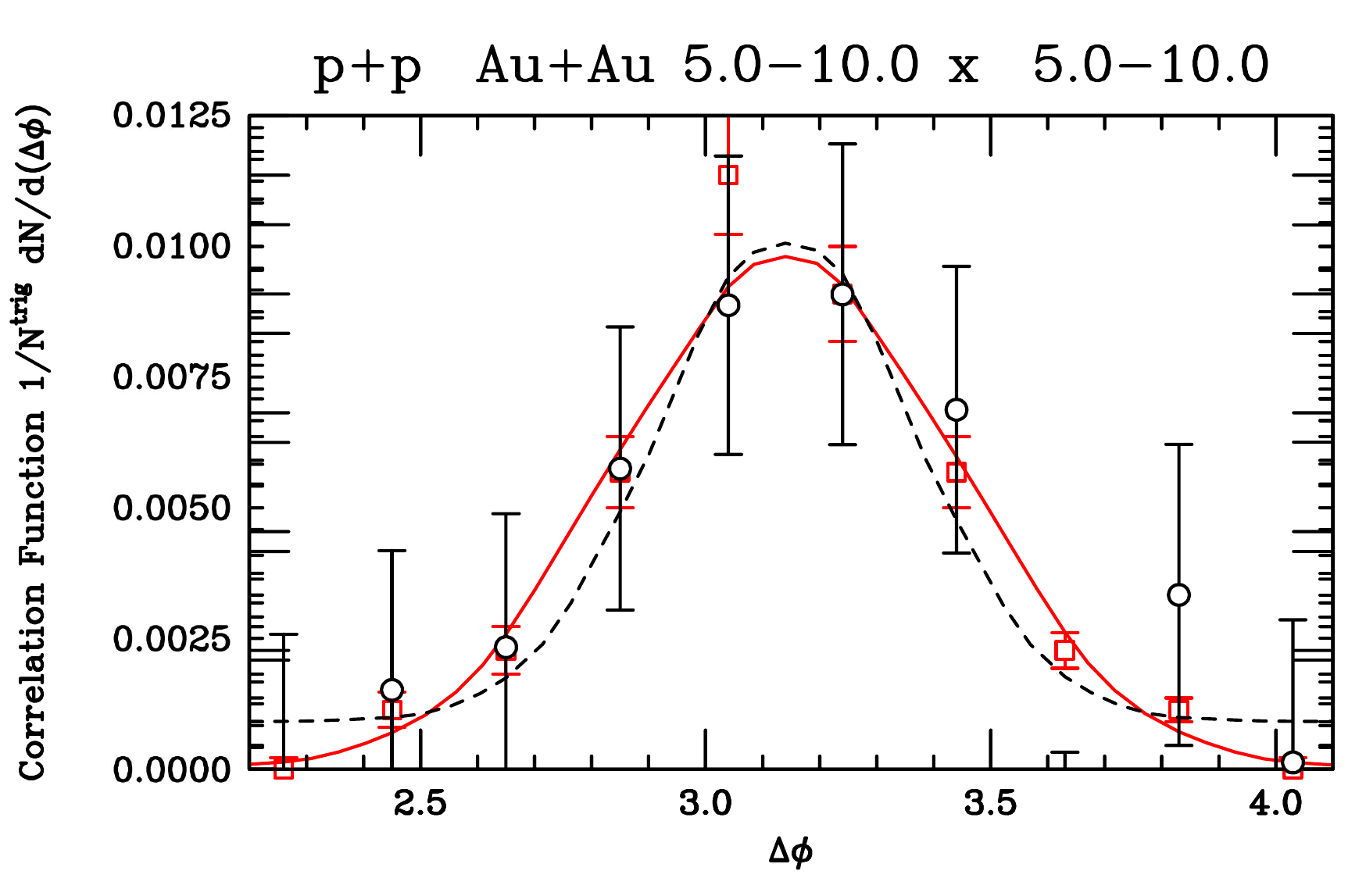}
\includegraphics[width=5.55cm]{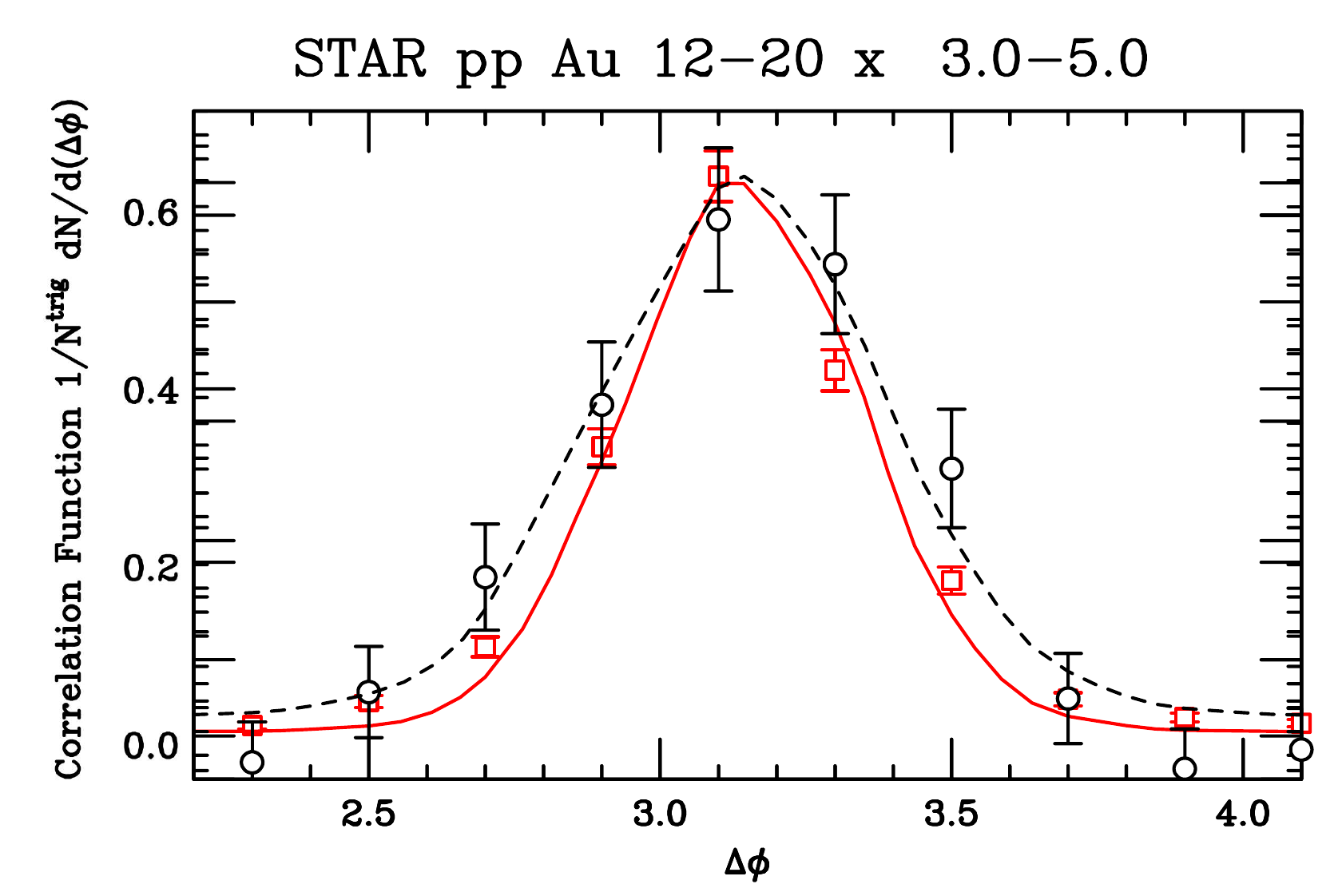}
\end{center}
\caption[*]{Gaussian fits to actual dihadron angular correlation measurements of PHENIX \cite{PXppg074} plus the previous fit~\cite{MJTPLB771} to the STAR data \cite{STARPLB760}. (p$+$p data open squares, fits solid lines; Au+Au data open circles, fits dashed lines). The y axes for the Au$+$Au data and fits are rescaled so that the peaks in the p$+$p and Au$+$Au fits lie on top of each other.}
\label{fig:PXdihadron}
\end{figure*}

In my opinion the derived PHENIX data in Fig.~\ref{fig:Bowendihadron} reproduced from Ref.\cite{ChenCCNUPLB773} looked nothing like the published PHENIX data in Fig.~\ref{fig:ppg074gh}. Admittedly a listing of the data in Fig.~\ref{fig:ppg074gh} was not available, but a following publication with the exact same figure~\cite{PXppg083} did provide a listing of the data points~\cite{ppg083data}~\footnote{Ref.~\cite{ppg083data} shows that these are the actual data from Fig.~1 of Ref~\cite{PXppg074}}. Given these actual data points for the PHENIX dihadron correlations shown in Fig.~\ref{fig:PXdihadron} I first fit the data to  Gaussians in $\Delta\phi$ ($\sigma_{\Delta\phi}$) for the away side $\pi/2\leq \Delta\phi\leq 3\pi/2$ and the trigger side $-\pi/2\leq \Delta\phi\leq\pi/2$ in order to compare the data and fits to Fig.~\ref{fig:Bowendihadron}. The y axis for the Au$+$Au data and fits in Fig.~\ref{fig:PXdihadron} are rescaled so that the peaks in the p$+$p and Au$+$Au fits lie on top of each other. The STAR data and fits from Ref.~\cite{MJTPLB771} are also shown in Fig.~\ref{fig:PXdihadron}.


The most notable observation about the fits in Fig.~\ref{fig:PXdihadron} is that for both $p_{Ta}$ ranges, the PHENIX Au$+$Au fits have smaller $\sigma_{\Delta\phi}$ than the p$+$p fits, which is more convenient to quote in the variable $\mean{p^2_{\rm out}}=(p_{Ta}\sin\sigma_{\Delta\phi})^2$ as follows: for PHENIX $p_{Tt}=5-10$ GeV/c,  
the values of $\mean{p^2_{\rm out}}$ for $p_{Ta}=3-5$ GeV/c are $0.79\pm 0.64$ (GeV/c)$^2$, $\chi^2$/dof=22/23,  for Au$+$Au 0-20\%  and $1.54\pm 0.08$ (GeV/c)$^2$ for p$+$p;  and for $p_{Ta}=5-10$ GeV/c,  $2.12\pm1.13$ (GeV/c)$^2$, $\chi^2$/dof=13/23, for Au$+$Au and $3.92\pm 0.33$ (GeV/c)$^2$ for p$+$p. For the STAR Au$+$Au 00-12\%, $p_{Tt}=12-20$, $p_{Ta}=3-5$ GeV/c data, the results are the same as in Ref.~\cite{MJTPLB771}, namely  $\mean{p^2_{\rm out}}=0.851\pm 0.203$ (GeV/c)$^2$ for Au$+$Au and $0.576\pm0.167$ (GeV/c)$^2$ for p$+$p.

From these numbers it is obvious~\cite{MJTPLB771} that $\mean{\hat{q}L}$ (which corresponds to the $\mean{p^2_{\perp}}$ on Fig.~\ref{fig:Bowendihadron}) is negative for the PHENIX data and thus not equal to $\mean{\hat{q}L}= 13$ GeV$^2$ quoted on Fig.~\ref{fig:Bowendihadron} reproduced from Ref.~\cite{ChenCCNUPLB773}.
For readers who may not understand this as obvious, a review of the method to calculate $\mean{\hat{q}L}$ is presented followed by the calculations of $\mean{\hat{q}L}$ from the PHENIX and STAR data in Fig.~\ref{fig:PXdihadron} and some other published PHENIX data, leading to an interesting conclusion.   
         
\section{A review and improvement of the method to measure $\mean{\hat{q}L}$ from di-hadron azimuthal broadening}
The BDMPSZ~\cite{BSZARNPS50} QCD based prediction for detecting the QGP is jet quenching produced by the energy loss, via LPM coherent radiation of gluons, radiated from  an outgoing parton with color charge fully exposed in a medium with a large density of similarly exposed color charges (i.e. the QGP). 
As a parton from hard-scattering in the A+B collision exits through the medium it can radiate a gluon; and both continue traversing the medium. It is important to understand that ``Only the gluons radiated outside the cone defining the jet contribute to the energy loss.''~\cite{BSZARNPS50}. Also because of the angular ordering of QCD ~\cite{MuellerPLB104}, the angular cone of any further emission will be restricted to be less than that of the previous emission and will end the energy loss once inside the jet cone. However complications might occur in the deconfined QGP~\cite{YacineIJMPA}.

The energy loss of the original outgoing parton, $-dE/dx$,  
per unit length ($x$) of a medium with total length $L$, is proportional to the total 4-momentum transfer-squared, $q^2(L)$, with the form:\vspace*{-0.5pc}
\begin{equation}{-dE \over dx }\simeq \alpha_s \langle{q^2(L)}\rangle=\alpha_s\, \mu^2\, L/\lambda_{\rm mfp} 
=\alpha_s\, \hat{q}\, L\qquad \label{eq:dEdx} \end{equation}
where $\mu$, is the mean momentum transfer per collision, and the transport coefficient 
{$\hat{q}=\mu^2/\lambda_{\rm mfp}$} is the 4-momentum-transfer-squared to the medium per mean free path, $\lambda_{\rm mfp}$.\\[0.5pc] 
Also, the accumulated momentum-squared, $\mean{p^2_{\perp W}}$ transverse to the  parton from its collisions traversing a length $L$ in the medium is well approximated by 
\begin{equation} 
\mean{p^2_{\perp W}}\approx\langle{q^2(L)}\rangle=\hat{q}\, L. \label{eq:pperpsq}
\end{equation}
This is strongly correlated to the energy loss~Eq.~\ref{eq:dEdx}~\cite{BDMPSNPB484} and results in the azimuthal broadening of the outgoing parton from its original direction by $\mean{\hat{q} L}/2$ since only the component $\mean{p^2_{\perp W}}/2$ is in the azimuthal direction, $\perp$ to the scattering plane. 
The original parton that scattered had a so-called~\cite{FFF} intrinsic mean transverse momentum $\mean{k_T}$ which is perpendicular to the collision axis but can act both perpendicular to the scattering plane and in the scattering plane at random.
This means that in a p$+$p collision, the mid-rapidity di-jets from hard-parton-parton scattering are not back-to-back in azimuth but are acollinear from the random sum of $\mean{k_T^2}$ from both scattered partons or $\mean{p_T^2}_{\rm pair}=2\mean{k_T^2}$, of which only half or $\mean{k_T^2}$ affects the azimuthal broadening  while the other half unbalances the original equal and opposite transverse momenta of the jets ~\cite{MJTPLB771}. 
 In an A$+$A collision this di-jet gets further broadened in azimuth by the random sum of the azimuthal component $\mean{p^2_{\perp W}}/2$ {\bf from each outgoing jet} or $\mean{p^2_{\perp W}}=\hat{q}\, L$, so that the di-jet azimuthal broadening acoplanarity in A$+$A collisions compared to p$+$p collisions should be
\begin{equation}  
\mean{\hat{q} L}=\mean{k_{T}^2}_{AA}-\mean{k{'}_{T}^2}_{pp} \label{eq:qhatL}
\end{equation}
 since only the component of $\mean{p^2_{\perp W}}$ $\perp$ to the scattering plane affects $k_T$. \footnotemark[2]~\footnotetext[2]{Ref.~\cite{MJTPLB771} had $\mean{\hat{q} L}/2=$ in Eq.~\ref{eq:qhatL} because I forgot that the di-hadron correlation represents both the trigger and away-side scattered partons.} This is the azimuthal di-jet broadening from the BDMPSZ energy loss in the medium. Here  $\mean{k_{T}^2}_{AA}$ denotes the intrinsic rms. transverse momentum of the hard-scattered parton in a nucleon in an A$+$A collision plus any medium effect; and $\mean{k{'}_{T}^2}_{pp}$ denotes the reduced value of the p$+$p comparison di-hadron $\mean{k_{T}^2}_{pp}$ measurement with $p_{Tt}$ and $p_{Ta}$ correcting for the lost energy of the scattered partons in the QGP~\cite{MJTPLB771}.
This reduces to the simpler equation when the equation for the $\mean{k^2_T}$ for di-hadrons is substituted~\cite{MJTPLB771}:
\begin{equation}
\mean{\hat{q} L}=\left[\frac{\hat{x}_h}{\mean{z_t}}\right]^2_{AA} \;\left[\frac{\mean{p^2_{\rm out}}_{AA} - \mean{p^2_{\rm out}}_{pp}}{x_h^2}\right] \label{eq:coolqhat}
\end{equation}
where $\mean{p^2_{\rm out}}={p}_{Ta}^2 \mean{\sin^2(\pi-\Delta\phi)}$ and the di-hadrons with $p_{Tt}$ and $p_{Ta}$, with ratio $x_h=p_{Ta}/p_{Tt}$, are assumed to be fragments of jets with transverse momenta $\hat{p}_{Tt}$ and $\hat{p}_{Ta}$ with ratio $\hat{x}_h=\hat{p}_{Ta}/\hat{p}_{Tt}$, where ${z_t}\simeq p_{Tt}/\hat{p}_{Tt}$ is the fragmentation variable, the fraction of momentum of the trigger particle in the trigger jet. 
For di-jet measurements, Eq.~\ref{eq:coolqhat} becomes even simpler: i) $x_h \equiv \hat{x}_h$ because the trigger and away `particles' are the jets; ii) $\mean{z_t}\equiv 1$ because the trigger `particle' is the entire jet not a fragment of the  jet; iii) $\mean{p^2_{\rm out}}=\hat{p}_{Ta}^2 \mean{\sin^2(\pi-\Delta\phi)}$. This reduces Eq.~\ref{eq:coolqhat} for di-jets to:
\begin{equation}
\mean{\hat{q} L}= \left[{\mean{p^2_{\rm out}}_{AA} - \mean{p^2_{\rm out}}_{pp}}\right]
\label{eq:jetqhat}
\end{equation}
I checked Eq.~\ref{eq:jetqhat} against a prediction~\cite{MuellerPLB763} for 35 GeV/c jets at RHIC at \sqsn=200 GeV with $\mean{\hat{q} L}$=0 GeV$^2$ (p$+$p), and for A$+$A, $\mean{\hat{q} L}$=8 GeV$^2$ and 20 GeV$^2$. I got 9.7 GeV$^2$ and 21.5 GeV$^2$ respectively for the 8 GeV$^2$ and 20 GeV$^2$ curves subtracting the p$+$p value of   
$\mean{p^2_{\rm out}}_{pp}$.
\subsection{How to Find $\mean{z_t}$, $\hat{x}_h$, and the energy loss of $\hat{p}_{Tt}$ for dihadrons}
\subsubsection{$\mean{z_t}$ and $\Delta\hat{p}_{Tt}$}
\begin{figure}[!hbt]
\includegraphics[width=0.85\linewidth]{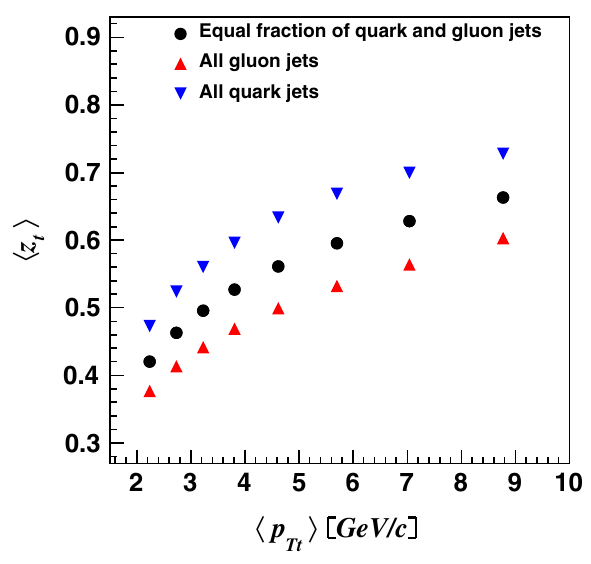}
\caption[]{Values of $\mean{z_t}$ as a function of $p_{Tt}$ for pure quarks, pure gluons and half and half~\cite{ppg089}. } 
\label{fig:meanz_t}
\end{figure}

At RHIC, in p$+$p and Au$+$Au collisions as a function of centrality the $\pi^0$ $p_{T}$ spectra with \mbox{$5 <p_T \ \lsim \ 20$ GeV/c} all follow the same power law with $n\approx 8.10\pm0.05$~\cite{ppg080}. 
The Bjorken parent-child relation and `trigger-bias'~\cite{JacobPLC48} then imply that the single particle cross section has the same power law shape, $d^3\sigma/2\pi p_T dp_T dy\propto p_T^{-n}$, as the parent jet  cross section and that large values of $\mean{z_t}=p_{Tt}/\hat{p}_{Tt}$ dominate the single-particle cross section. This means that the shift ($\Delta\hat{p}_{Tt}$) in the A$+$A Jet $p_T$ spectrum at a given $p_{Tt}$ from the $\mean{T_{AA}}$ corrected p$+$p cross section can be measured from the shift in the trigger hadron spectrum~\cite{ppg133,MJTPLB771}. Similarly, the $\mean{z_t}$ as a function of $p_{Tt}$ can be calculated~\cite{ppg089,ppg029} (Fig.~\ref{fig:meanz_t}) using the measured fragmentation functions from e$+$p collisions~\cite{DELPHI,OPAL}. The difference in $\mean{z_t}$ for p$+$p and Au$+$Au due to the shift in the spectrum is considerably less than the error in the calculated $\mean{z_t}$ so the $\mean{z_t}$ calculated from p$+$p spectrum  is used for the Au$+$Au spectrum with the same $p_{Tt}$~\cite{MJTPLB771}.

\subsubsection{The $x_E\approx p_{Ta}/p_{Tt}$ distribution from a $p_{Tt}$ trigger measures the ratio of the away jet to the trigger jet $\hat{p}_{T}$: $\hat{x}_h=\hat{p}_{Ta}/\hat{p}_{Tt}$.}
As discussed in Ref.~\cite{MJTPLB771} it was assumed since the high $p_T$ discovery at the CERN ISR that the \mbox{$x_E=\cos{\Delta\phi}\times p_{Ta}/p_{Tt}\approx x_h$} disribution would measure the away-jet fragmentation function as is does for direct-photon triggers~\cite{ppg095}. However it was found at RHIC~\cite{ppg029} that the $x_E$ distribution (which PHENIX calls $x_h$ and STAR calls $z_T$) measured the ratio of the away-jet to trigger-jet transverse momenta 
$\hat{x}_h=\hat{p}_{Ta}/\hat{p}_{Tt}$ (Eq.~\ref{eq:condxe2N})
    \begin{equation}
\left.{dP_{\pi} \over dx_E}\right|_{p_{T_t}}  = {N\,(n-1)}{1\over\hat{x}_h} {1\over {(1+ {x_E \over{\hat{x}_h}})^{n}}} \,  
\qquad ,  
\label{eq:condxe2N}
\end{equation}   
with the value of $n=8.10$ ($\pm0.05$) fixed as determined in  Ref.~\cite{ppg080}, where $n$ is the power-law of the inclusive $\pi^0$ spectrum and is observed to be the same in p$+$p and Au$+$Au collisions in the $p_{T_t}$ range of interest. 

\begin{figure}[!th] 
\begin{center}
\raisebox{0.0pc}{\includegraphics[width=0.85\linewidth]{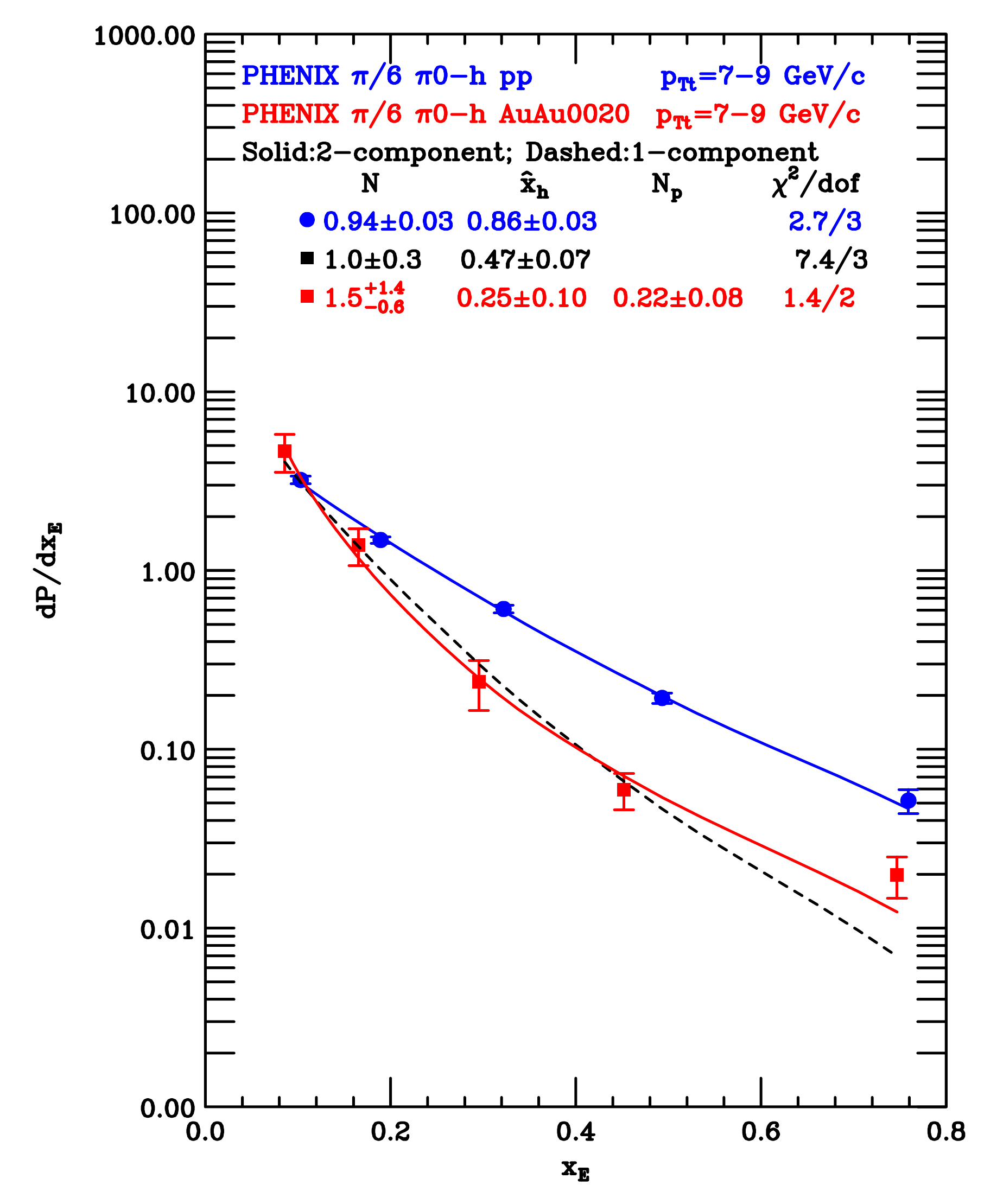}}
\end{center}\vspace*{-1.0pc}
\caption[]{Fits to PHENIX $d{\mathcal{P}}/dx_E$ distributions~\cite{ppg106,MJTGrenoble} for $\pi^0$-h correlations with $7\leq p_{Tt}\leq 9$ GeV/c in \sqsn=200 GeV p$+$p and Au$+$Au 0-20\% collisions.}  
\label{fig:hatxh}
\end{figure}

Figure~\ref{fig:hatxh} shows a fit of Eq.~\ref{eq:condxe2N} to the PHENIX $x_E$ Au$+$Au \hbox{0-20\%} and p$+$p distributions in a region with $\mean{p_{Tt}}\approx 7.8$ GeV/c, close to the $5\leq p_{Tt}<10$ GeV/c region in Fig.~\ref{fig:PXdihadron} with $\mean{p_{Tt}}\approx 6.5$ GeV/c. The results are $\hat{x}_h=0.86\pm 0.03$ in p$+$p and $\hat{x}_h=0.47\pm 0.07$ Au$+$Au (dashes). What is more interesting is a fit to Eq.~\ref{eq:condxe2N} for $N$ and $\hat{x}_h$ plus another term of Eq.~\ref{eq:condxe2N} with $\hat{x}_h=0.86$ fixed at the p$+$p value, with the normalization $N_p=0.22\pm 0.08$ fitted, compared to the $N=1.5^{+1.4}_{-0.6}$ for the partons that have lost energy.  The result is the solid Au$+$Au curve with a much better $\chi^2$ which is notably parallel to the p$+$p curve for $x_E\geq 0.4$ ($p_{Ta}\approx p_{Tt}\times x_E= 3.1$ GeV/c).
\vspace*{-2.0pc}

\subsubsection{This effect is well known under a different name}
One possible explanation is that in this region for \mbox{$p_{Ta}\geq 3$ GeV/c}, which is at a fraction $\approx 1\%$ of the $\left.{dP_{\pi}/dx_E}\right|_{p_{T_a}}$ distribution, 
these hard fragments are distributed narrowly around the jet axis so that they are not strongly affected by the medium~\cite{YacineIJMPA}. 
An unlikely possibility is from tangential parton-parton collisions at the periphery of the A$+$A overlap region which has probability much smaller than the $N_p/N$ ratio.  

\begin{figure}[]
\begin{center}
\raisebox{0.0pc}{\includegraphics[width=0.95\linewidth]{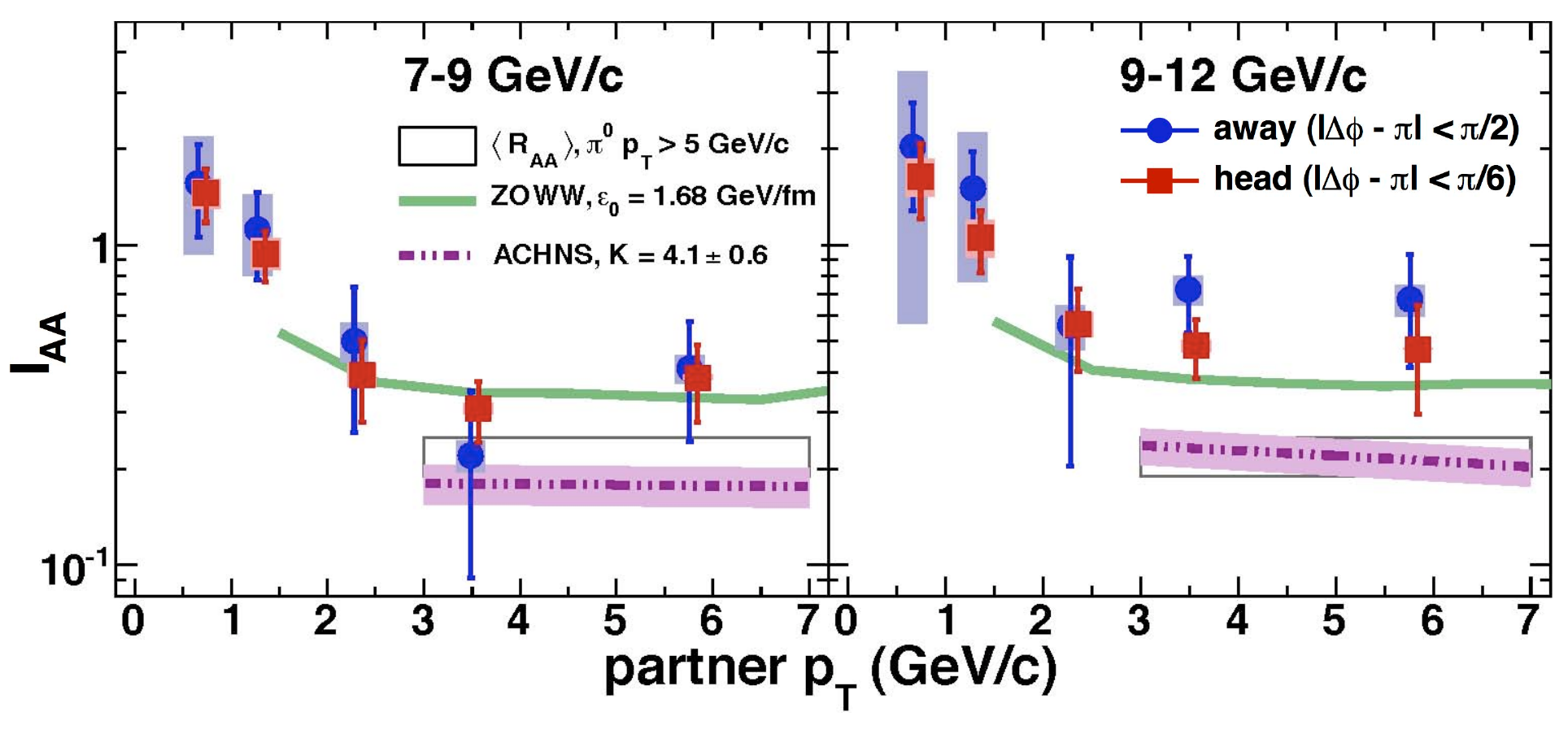}}
\end{center}\vspace*{-1.0pc}
\caption[]{PHENIX~\cite{ppg106}  $I_{AA}=d{\mathcal{P}}/dx_E|_{AA}/d{\mathcal{P}}/dx_E|_{pp}$  for $p_{Tt}$=7-9 and 9-12 GeV/c vs. partner $p_T$ (i.e. $p_{Ta}$) in \sqsn=200 GeV p$+$p and Au$+$Au 0-20\% collisions} 
\label{fig:ppg106IAA}
\end{figure}

Either possibility is consistent with measurements of the ratio of the Au$+$Au to p$+$p  $x_E$ (or $p_{Ta}$) distributions for a given $p_{Tt}$ which are called $I_{AA}$ distributions (Fig.~\ref{fig:ppg106IAA}~\cite{ppg106}). 
All $I_{AA}$ distributions ever measured show the same effect as in Fig.~\ref{fig:ppg106IAA}, they fall in the range $0<p_{Ta}<3$ GeV/c and then remain constant.  The same effect can be seen in an $I_{AA}$ measurement in \sqsn=2.76 TeV p$+$p and Pb$+$Pb 0-10\% by ALICE at the LHC~\cite{ALICEPLB763}. The fact that $I_{AA}$ remains constant above $p_{Ta}\approx 3$ GeV/c means that the ratio of the away-jet to the trigger jet transverse momenta in this region remains equal in A$+$A and p$+$p , i.e. no apparent suppression via energy loss in this region. This effect also causes problems in the following calculations of $\mean{\hat{q} L}$ from the di-hadron 
correlations.

\section{Calculation of $\mean{\hat{q} L}$ from di-hadron azimuthal broadening.}
The calculations of $\mean{\hat{q} L}$ for the STAR measurement~\cite{STARPLB760} in Fig.~\ref{fig:PXdihadron} as well as for $1.0\leq p_{Ta}\leq 3$ GeV/c performed in Ref~\cite{MJTPLB771}~\footnotemark[3]~\footnotetext[3]{The sharp-eyed reader will notice that the $\mean{\hat{q}L}$ values in Ref.~\cite{MJTPLB771} were $8.41\pm2.66$  and $1.71\pm 0.67$ GeV$^2$ for two reasons: first is the $\mean{\hat{q}L}/2$ in Eq.~\ref{eq:qhatL} there, second was a miscalculation of the error which should have been obvious from the errors in $\mean{p^2_{\rm out}}$ which are unchanged.} with the values $\hat{x}^{pp}_h=0.84 \pm 0.04$, $\mean{z_t}=0.80\pm 0.05$ are given in Table~\ref{tab:star-PLB760}.
   \begin{table}[!h]\vspace*{-0.0pc} \label{tab:star-PLB760}
\begin{center}
\caption[]{Tabulations for $\hat{q}$--STAR $\pi^0$-h~\cite{STARPLB760}} 
{\begin{tabular}{ccccc} 
\hline
\hline
STAR PLB760\\
\hline
$\sqsn=200$ &$\mean{p_{Tt}}$ & $\mean{p_{Ta}}$ & $\mean{p^2_{\rm out}}$& $\mean{\hat{q} L}$ \\ 
\hline
Reaction & GeV/c &  GeV/c & (GeV/c)$^2$& GeV$^2$ \\ 
\hline
p$+$p&14.71&1.72&$0.263\pm 0.113$ & \\
\hline
p$+$p&14.71&3.75&$0.576\pm 0.167$&\\
\hline
Au$+$Au 0-12\%&14.71&1.72&$0.547\pm 0.163$&$4.21\pm3.24$\\
\hline
Au$+$Au 0-12\%&14.71 &3.75&$0.851\pm 0.203$&$0.86\pm 0.87$\\
\hline\\[-0.9pc]
\hline
\end{tabular}} \label{tab:star-PLB760}
\end{center}\vspace*{-0.1pc}
\end{table}
The value of $\mean{\hat{q} L}=0.86\pm 0.87$ GeV$^2$ for the fit to the $3\leq p_{Ta}\leq 5$ GeV/c STAR data shown in Fig.~\ref{fig:PXdihadron} is consistent with zero and clearly in significant disagreement with the proposed $\mean{\hat{q} L}=\mean{p^2_{\perp}}=13$ GeV$^2$ quoted on Fig.~\ref{fig:Bowendihadron}~\cite{ChenCCNUPLB773}. The value of $\mean{\hat{q} L}=4.21\pm 3.24$ GeV$^2$ in the lower $p_{Ta}$ bin is closer to the prediction, within 2.7 standard deviations, but also consistent with zero. 

The calculations of $\mean{\hat{q} L}$ from the fits to the PHENIX data in Fig.~\ref{fig:PXdihadron} with $\hat{x}_h=0.51\pm 0.06$ and $\mean{z_t}=0.64\pm 0.64$ are given in Table~\ref{tab:PHENIXppg083}. 
The values of $\mean{\hat{q} L}=-2.24\pm2.01$ and $-1.68\pm 1.20$ GeV$^2$ are negative, as noted above, and both consistent with zero but inconsistent with the predicted 13 GeV$^2$.

   \begin{table}[!h]\vspace*{-0.0pc} \label{tab:PHENIXppg083}
\begin{center}
\caption[]{Tabulations for $\mean{\hat{q}L}$--PHENIX Fig.~\ref{fig:PXdihadron}} 
{\begin{tabular}{ccccc} 
\hline
\hline
PHENIX PRC77\\
\hline
$\sqsn=200$ &$\mean{p_{Tt}}$ & $\mean{p_{Ta}}$ & $\mean{p^2_{\rm out}}$& $\mean{\hat{q} L}$ \\ 
\hline
Reaction & GeV/c &  GeV/c & (GeV/c)$^2$& GeV$^2$ \\ 
\hline
p$+$p&8.08&3.75 &$1.54\pm 0.08$ &\\
\hline
p$+$p&8.08&6.68&$3.92\pm 0.33$&\\
\hline
Au$+$Au 0-20\%&8.08&3.75 &$0.79\pm 0.64$ & $-2.24\pm2.01$ \\
\hline
Au$+$Au 0-20\%&8.08 &6.68&$2.12\pm 1.13$&$-1.68\pm 1.21$\\
\hline\\[-0.9pc]
\hline
\end{tabular}} \label{tab:PHENIXppg083}
\end{center}\vspace*{-1.1pc}
\end{table}

\begin{figure}[!h]
\begin{center}
\raisebox{0.0pc}{\includegraphics[width=0.95\linewidth]{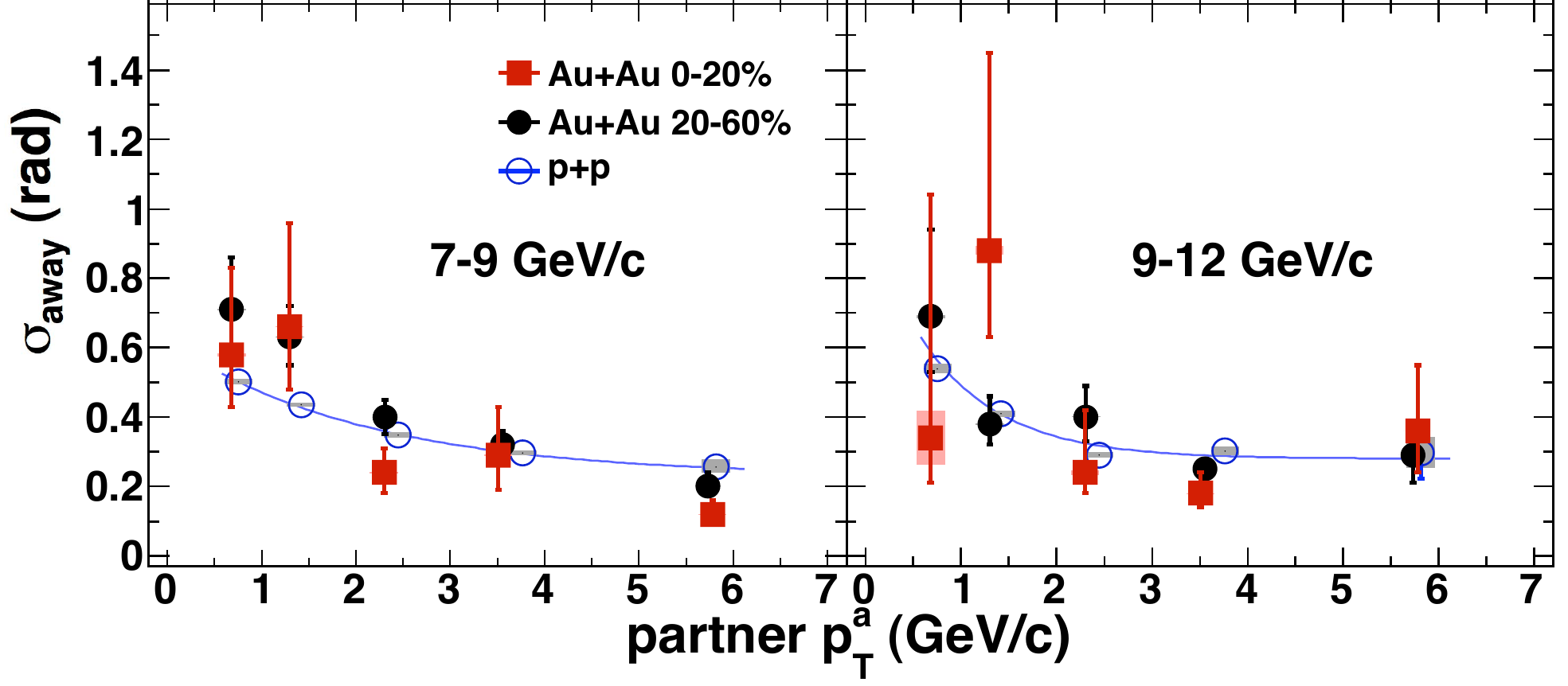}}
\end{center}
\caption[]{PHENIX~\cite{ppg106}  $\sigma_{\rm away}$ for $p_{Tt}$=7-9 and 9-12 GeV/c vs. partner $p_T\equiv p_{Ta}$} 
\label{fig:ppg106sigma}
\end{figure}

Although not discussed in Ref.~\cite{ChenCCNUPLB773}, the PHENIX measurement of $I_{AA}$ shown in Fig.~\ref{fig:ppg106IAA} also provided values of $\sigma_{\rm away}$ for Au$+$Au and p$+$p  plotted clearly (Fig.~\ref{fig:ppg106sigma}) so that values of $\hat{q}L$ can be read off practically by inspection. While $\sigma_{\rm away}$ is apparently larger in Au$+$Au than in p$+$p for $p_{Ta}<2$ GeV/c it is smaller or equal to the p$+$p value for $p_{Ta}>2$ GeV/c, i.e. $\hat{q}L$ consistent with zero. Details for $p_{Tt}=9-12$ GeV/c are given in Table \ref{tab:PHENIXppg106}. 

   \begin{table}[t]\vspace*{-0.0pc} \label{tab:PHENIXppg106}
\begin{center}
\caption[]{Tabulations of $\mean{\hat{q}L}$-PHENIX 9-12 GeV/c Fig.~\ref{fig:ppg106sigma}} 
{\begin{tabular}{ccccc} 
\hline
\hline
PHENIX PRL104\\
\hline
$\sqsn=200$ &$\mean{p_{Tt}}$ & $\mean{p_{Ta}}$ & $\mean{p^2_{\rm out}}$& $\mean{\hat{q} L}$ \\ 
\hline
Reaction & GeV/c &  GeV/c & (GeV/c)$^2$& GeV$^2$ \\ 
\hline
p$+$p&10.22&1.30&$0.319\pm 0.023$ &\\
\hline
p$+$p&10.22&2.31&$0.491\pm 0.052$&\\
\hline
p$+$p&10.22&3.55&$1.256\pm 0.166$ &\\
\hline
p$+$p&10.22&5.73&$2.884\pm 1.376$&\\
\hline
Au$+$Au 0-20\%&10.22&1.30 &$0.86\pm 0.339$ & $13.3\pm10.4$ \\
\hline
Au$+$Au 0-20\%&10.22&2.31&$0.299\pm0.190$&$-1.5\pm1.7$\\
\hline
Au$+$Au 0-20\%&10.22&3.55&$0.394\pm 0.189$ & $-2.9\pm1.6$ \\
\hline
Au$+$Au 0-20\%&10.22&5.73&$4.08\pm2.83$&$ 1.5\pm 4.0$\\
\hline\\[-0.9pc]
\hline
\end{tabular}} \label{tab:PHENIXppg106}
\end{center}\vspace*{-1.0pc}
\end{table}

\section{Conclusion} When calculated with fits to the measured distributions in Fig.~\ref{fig:PXdihadron} the values of $\hat{q}L$ are inconsistent with the calculation of $\hat{q}L=13$ GeV$^2$ claimed in Fig.~\ref{fig:Bowendihadron}~\cite{ChenCCNUPLB773}, for $p_{Ta}\geq 3$ GeV/c. For values of $p_{Ta}<3$ GeV/c, separating the flow background causes the errors in the measurement of $\hat{q}L$ to be too large to obtain a reasonable value.
 
The measurement of $\hat{q}L$ and possibly $\hat{q}$ can be greatly improved by measuring di-jet angular distributions rather than di-hadron distributions. The energy loss of the trigger jets can be determined by the shift in the $p_{Tt}$ spectrum from p$+$p to A$+$A the same way as for $\pi^0$~\cite{ppg133,MJTPLB771}. Then a plot of the $\hat{p}_{Ta}$ of the away jets for a given trigger jet with $\hat{p}_{Tt}$ analogous to Fig.~\ref{fig:hatxh} and an evaluation of $\Delta E=\alpha_s\, \hat{q}\, L^2$ from $\hat{p}_{Tt}-\hat{p}_{Ta}$ and $\hat{q}L$ by Eq.~\ref{eq:jetqhat} as a function of $\hat{p}_{Ta}$  might allow the $L$ dependence to be factored out or determined which would lead to a experimental measurement of $\hat{q}$.

\begin{acknowledgments}   
Research supported by U.~S.~Department of Energy, Contract No. {DE-SC0012704}.
\end{acknowledgments}

\end{document}